# GRAPH-BASED PYRAMID GLOBAL CONTEXT REASONING WITH A SALIENCY-AWARE PROJECTION FOR COVID-19 LUNG INFECTIONS SEGMENTATION

*Huimin Huang[1], \*Ming Cai[1], \*Lanfen Lin[1], Jing Zheng[2], Xiongwei Mao[2], Xiaohan Qian[2], Zhiyi Peng[2], \*Jianying Zhou[2], Yutaro Iwamoto[3], Xian-Hua Han[3], \*Yen-Wei Chen[3,4,1], Ruofeng Tong[1]*

[1] College of Computer Science and Technology, Zhejiang University, China
[2] College of Medicine, The First Affiliated Hospital, Zhejiang University, China
[3] College of Information Science and Engineering, Ritsumeikan University, Japan
[4] Research Center for Healthcare Data Science, Zhejiang Lab, Hangzhou, China

## ABSTRACT

Coronavirus Disease 2019 (COVID-19) has rapidly spread in 2020, emerging a mass of studies for lung infection segmentation from CT images. Though many methods have been proposed for this issue, it is a challenging task because of infections of various size appearing in different lobe zones. To tackle these issues, we propose a Graph-based Pyramid Global Context Reasoning (Graph-PGCR) module, which is capable of modeling long-range dependencies among disjoint infections as well as adapt size variation. We first incorporate graph convolution to exploit long-term contextual information from multiple lobe zones. Different from previous average pooling or maximum object probability, we propose a saliency-aware projection mechanism to pick up infection-related pixels as a set of graph nodes. After graph reasoning, the relation-aware features are reversed back to the original coordinate space for the down-stream tasks. We further construct multiple graphs with different sampling rates to handle the size variation problem. To this end, distinct multi-scale long-range contextual patterns can be captured. Our Graph-PGCR module is plug-and-play, which can be integrated into any architecture to improve its performance. Experiments demonstrated that the proposed method consistently boost the performance of state-of-the-art backbone architectures on both of public and our private COVID-19 datasets.

## 1. INTRODUCTION

The break of Coronavirus Disease 2019 (COVID-19) has rapidly spread over the world, which has been declared as a global pandemic [1]. Accurate lung infections segmentation is one of the most important pre-processing steps for assessment and quantification of COVID-19 [2-5]. The classic UNet [6] and UNet++ [7] were widely performed as segmentation architectures for COVID-19. Recently, UNet-Inf [8] with a parallel partial decoder was proposed to segment lung infections. Despite achieving good results, these approaches are still incapable of exploring sufficient information from multifocal infections, appearing in different lobe zones [9,10]. It may hinder the infections segmentation performance, especially considering each pixel in isolation, as local information is noisy and ambiguous. It is also noteworthy that infections with various size occur in different scales. To tackle these two challenges, it is imperative to perform multi-scale long-term interactions on COVID-19 CT images, which contributes to model long-term dependencies among multiple lesions.

Recently, graph convolution [11] has been incorporated into computer vision tasks for globally reasoning, which can be generally summarized as two kinds of approaches: feature space graph convolution and coordinate space graph convolution. The feature space graph convolution captures interdependencies along the channel dimensions of the feature map, which projects the feature into a non-coordinate space [12-15]; whistle coordinate space graph convolution explicitly models the spatial relationships between pixels [16-20], which projects the feature into a new coordinate space, to produce coherent prediction between the disjoint infections.

In this paper, we propose a saliency-aware projection-based **G**raph-based **P**yramid **G**lobal **C**ontext **R**easoning (Graph-PGCR) module for COVID-19 lung infections segmentation. Different from the existing work that the infection-related pixels were highlighted via average pooling [16] or maximum object probability [17], we propose a saliency-aware projection (SAP) to keep eye on 'where' is an informative part, and thus selects discriminative pixels to form a fully-connected graph. In addition, we further take the multi-scale cues into consideration to address the challenge that different infections appear in various scales. Inspired by the Pyramid Pooling Module [21], we build a pyramid global context reasoning architecture to harvest multi-scale representations via SAP with various sampling rates. Hence, a coarser graph is constructed with lower sampling rate, providing more global dependencies for the larger receptive scale; while a finer graph is modeled with higher sampling rate, embedding more explicit long-range context for the smaller receptive field. In this way, we can perform graph reasoning on each scale and aggregate local and global clues to make the final prediction more reliable.

Our Graph-PGCR module is plug-and-play and thus can be integrated into a wide variety of existing network architectures to further enhance their performance. In summary, the main contributions of this research are four-fold: (**i**) We propose a Graph-PGCR module to model long-range dependencies among disjoint infections as well as adapt size variation; (**ii**) We propose a SAP mechanism to select the infection-related pixels as a set of graph nodes, where global contextual information can be propagated via graph convolution; (**iii**) We construct multiple graphs to harvest multi-scale contextual patterns from infections with various size; (**iv**) We conduct extensive experiments on public and private COVID-19 dataset, where our method yields consistent improvements over a number of baselines.

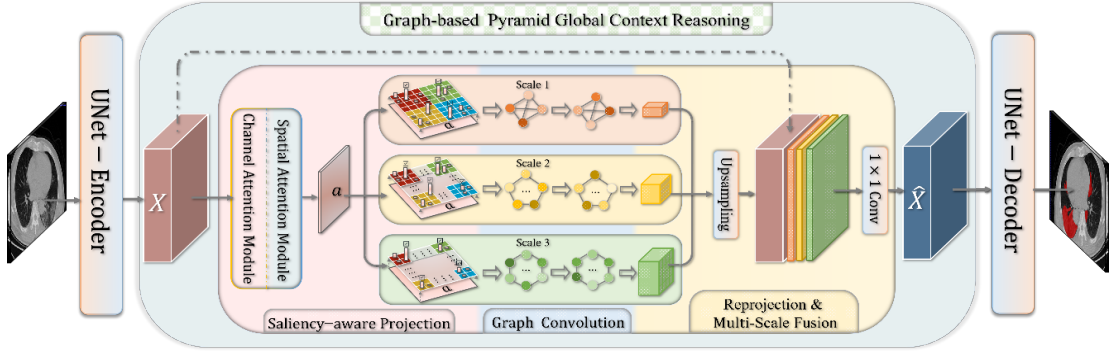

Fig. 1. Overview of the Graph-based Pyramid Global Context Reasoning (Graph-PGCR) module in UNet architecture.

## 2. METHODS

Fig.1 illustrates an overview of the proposed Graph-PGCR module in the segmentation architecture (e.g. UNet). Given an input image, we first extract features via the UNet-Encoder, and then our Graph-PGCR module is integrated to capture multi-scale long-range representations. Benefitting from the saliency-aware projection, the input feature map $X$ is firstly sampled into $K$ (e.g. $K = 3$) parallel pyramid levels with various scales. After individual graph convolution, the reprojection via upsampling and the multi-scale fusion with concatenation layers are performed to generate the feature representation $\hat{X}$, which is finally fed into the UNet-decoder for prediction. It is worth noting that the input feature map $X$ can be extracted from any layer of deep convolutional model. In the following subsections, we introduce the detail of each component in the Graph-PGCR module.

### 2.1. Saliency-aware Projection in Coordinate Space

In order to project infection-related pixels in coordinate space into a set of graph nodes in a new coordinate space, we proposed a saliency-aware projection mechanism, which integrates attention mechanism with pooling operation. Specifically, the attention mechanism aims at learning where to emphasize or suppress; while the pooling operation desires of picking out discriminative pixels.

Before projection, we need to reduce the dimension of feature map $X$, which enhances the capacity of the projection. Inspired by the dual attention network [22], we implement the channel attention module to capture the channel dependencies between any two channel maps via self-attention mechanism. After enhancing the feature representation, we adopt a 1×1 convolution layer to reduce the feature dimension from $C$ to $S$. Similarly, the feature map also enhanced by spatial attention module to model the spatial dependencies between any two positions. Benefiting from the channel and spatial attention modules, we could emphasize interdependent feature maps and improve the feature representation of specific semantics.

Considering that pooling along the channel dimension can effectively highlight informative regions [23], we further perform the max-pooling and average-pooling operations on the channel axis and then concatenate them to generate an attention map $a \in \mathbb{R}^{H \times W}$ which focus on the salient pixels related to infections and surpass unnecessary ones. As vividly shown in Fig.2, the attention map $a$ with spatial size of $H \times W$ is divided into several non-overlapping sub-regions with the stride of $\delta$ pixels. Within each region, the pixel $\hat{p}$ with maximum localization probability: $\hat{p} = \arg\max_{p_i} a(p_i)$ is selected as a node. This process results in a set of nodes $\mathcal{N} = \{n_j\}_{j=1}^{|\mathcal{N}|}$ for the feature map $X$. $|\mathcal{N}|$ equals to $\lceil H/\delta \rceil \times \lceil W/\delta \rceil$ and represents the number of nodes. $\lceil \cdot \rceil$ is the ceiling operation, which gives the smallest integer equal or larger than its input. Note that the process can be considered as a sampling process and $1/\delta$ can be considered as a sampling rate. In view of this, each node $n_j$ is represented by its corresponding image coordinates. It is worth noting that the spatial interval between nodes can be controlled by adjusting $\delta$. The coarser graph is constructed with lager $\delta$ values, which perhaps captures longer-range interactions among nodes. In contrast, all pixels are assigned as individual nodes in the extreme case, where $\delta = 1$. The initial feature representation of each $n_j$ is extracted from the feature map $\tilde{X}$ enhanced in both channel and spatial dimensions. This results in a set of node features, $\mathcal{Z} \in \mathbb{R}^{S \times |\mathcal{N}|}$, where $S$ equals to the feature dimension of $\tilde{X}$.

### 2.2. Multi-Scale Reasoning with Graph Convolution

In spite of graph reasoning exploring the global context, the long-term context pattern differs in multiple scales of the same image. Specifically speaking, the finer representation with smaller receptive field (smaller $\delta$) embeds more explicit context; while the coarser representation with larger receptive scale (larger $\delta$) explores global dependencies. Taking the multi-scale schema into consideration, we incorporate it with graph reasoning to extend the long-range contextual patterns, and thus devise the Graph-PGCR module.

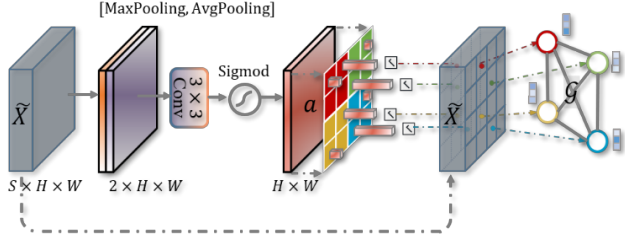
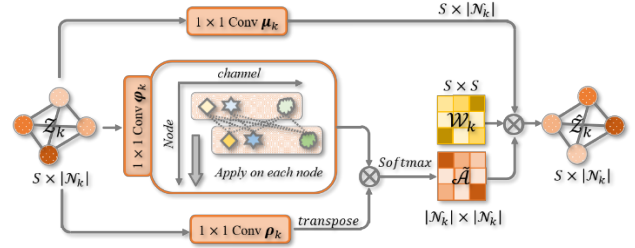

Fig. 2. An illustration of the proposed saliency-aware projection (SAP) mechanism. In this example, the attention map $a \in \mathbb{R}^{4 \times 4}$ is sampled with the stride $\delta = 2$. We then find the pixel $\hat{p}$ with maximum localization probability in each region (shown in different colors), which is selected as the node.

As seen in Fig.1, the graph convolution begins with subsampling the convolved features into $K$ parallel pyramid levels with various scales via saliency-aware projection. Higher sampling rate (smaller $\delta$) generates finer representations; while the coarser features are extracted from lower sampling rate (larger $\delta$). After selecting infection-related pixels as a set of nodes, a lightweight fully-connected graph with adjacency matrix $\mathcal{A}_k \in \mathbb{R}^{|\mathcal{N}_k| \times |\mathcal{N}_k|}$ is generated from $k$-th separate branch for propagating information across nodes, as depicted in Fig.3. The adjacency matrix $\mathcal{A}_k$ is defined as the similarity between nodes, where the more similar feature representations of two nodes, the stronger connectivity between them. It can be formulated as:

$$\mathcal{A}_k = \rho_k(\mathcal{Z}_k)^T \otimes \varphi_k(\mathcal{Z}_k) \qquad (1)$$

where $\rho_k(\cdot)$ and $\varphi_k(\cdot)$ are two learnable 1-dimensional linear transformations along node-wise dimension, $\otimes$ is the matrix multiplication. We further apply a softmax layer to yield a normalized adjacency matrix $\tilde{\mathcal{A}}_k$. Then we conduct the graph convolution [11] in our model as:

$$\tilde{\mathcal{Z}}_k = \sigma(\tilde{\mathcal{A}}_k (\mu_k(\mathcal{Z}_k))^T \mathcal{W}_k)^T \qquad (2)$$

where $\mu_k(\cdot)$ is a learnable linear transformation, $\mathcal{W}_k \in \mathbb{R}^{S \times S}$ is a trainable weight matrix, $\sigma(\cdot)$ is the ReLU activation function, and $\tilde{\mathcal{Z}}_k$ is the output feature map after graph convolution in $k$-th separate branch.

### 2.3. Reprojection and Multi-scale Fusion

To provide complementary feature for the down-stream task, the last step is to map the relation-aware features ($\mathbb{R}^{S \times |\mathcal{N}_k|}$) generated from $k$-th separate branch back to the coordinate space ($\mathbb{R}^{C \times H \times W}$), which is compatible with the regular CNN. To achieve the dimension transformation, we reshape the relation-aware $\tilde{\mathcal{Z}}_k \in \mathbb{R}^{S \times |\mathcal{N}_k|}$ into $\hat{\mathcal{Z}}_k \in \mathbb{R}^{S \times \lceil H/\delta_k \rceil \times \lceil W/\delta_k \rceil}$. Then, a simple but effective upsampling operation is adopted as the reprojection function. In practice, the bilinear interpolation is performed to resize $\lceil H/\delta_k \rceil \times \lceil W/\delta_k \rceil$ to the original spatial input size $H \times W$. To maintain the original information, we further utilize a multi-scale fusion to fuse the reshaped relation-aware features from each scale with the original feature map $X$ in a learnable way, which carries both local and global context information. The multi-scale fusion process can be formulated as:

Fig. 3. An illustration of reasoning with Graph convolution

$$\hat{X} = F\left(\left[\mathcal{U}(\tilde{\mathcal{Z}}_k)_{k=1}^K, X\right]\right) \qquad (3)$$

where $F(\cdot)$ realizes the feature aggregation mechanism with a 1×1 convolution followed by a batch normalization and a ReLU activation function. $\mathcal{U}(\cdot)$ indicate up-sampling operation, and $[\cdot]$ represents the concatenation. As a result, we have the feature $\hat{X}$ with channel dimension of $C$.

## 3. EXPERIMENTS AND RESULTS

### 3.1. Datasets and Implementation

The method was evaluated on two datasets: the public and our private COVID-19 datasets. (**i**) **The public COVID-19 dataset** [24]: It contains 20 COVID-19 CT scans from the Coronacases Initiative and Radiopaedia, which were manually annotated for the left lung, right lung and COVID-19 infection. In the experiment, we trained our models using the 16 volumes with 2-fold cross-validation and average the experiment results as the final performance. (**ii**) **The private COVID-19 dataset**: we collected 102 COVID-19 CT scans (from the Department of Radiology, The First Affiliated Hospital, College of Medicine, Zhejiang University), which has passed the ethic approvals. The left lung, right lung, and infection were annotated by two radiologists with 5-year experience in chest radiology. For our study, 82 scans were randomly selected for training and the other 20 scans for testing. After 2-fold cross-validation, we averaged the experiment results as the final performance.

The input image consisted of three slices: the slice to be segmented and the upper and lower slices, which was cropped to 224×224×3. The networks were updated utilizing the stochastic gradient descent, where the learning rate was 1e-3 and weight decay was 5e-4. To effectively model the global contextual information, our Graph-PGCR module was appended at the end of the encoder as seen in Fig.1. After feature extraction, the input feature map $X$ had the size of 1024×14×14. We simply set node feature dimension $S = 64$ in our implementation. The Dice coefficient was employed as our principal performance metric for each case.

### 3.1. Ablation Study

This section experiments the effect of key components of the Graph-PGCR module on the public COVID-19 dataset for lung infection segmentation. It includes the architecture of the original UNet as a baseline, Dual Attention (DA) Module [22], the proposed Graph-PGCR with different projection mechanisms and multi-scales with different $\delta$. Table 1 shows the segmentation performances when gradually adding components to the UNet. As seen, each component of the Graph-PGCR module contributes to the performance. Generally, a total improvement of 3.88% was gained by our proposed Graph-PGCR module compared to baseline UNet.

Table 2. Comparison with state-of-art methods on both public dataset ($Dice^{pu}$) and private dataset ($Dice^{pr}$).

| | Method | $Dice^{pu}$(%) | $Dice^{pr}$(%) | Method | $Dice^{pu}$(%) | $Dice^{pr}$(%) |
|---|---|---|---|---|---|---|
| UNet [6] | Baseline | 77.50 | 76.02 | UNet ++ [7] | 79.84 | 77.14 |
| | DA [22] | 78.11 | 77.06 | | 80.57 | 77.98 |
| | GloRe [13] | 78.23 | 76.95 | | 80.29 | 77.53 |
| | **Graph-PGCR ($\delta = 2$)** | **80.84** | **77.47** | | **81.03** | **78.60** |
| | **Graph-PGCR ($\delta = 2, 4, 7$)** | **81.38** | **78.92** | | **81.95** | **79.88** |
| UNet_Inf [8] | Baseline | 78.63 | 78.46 | UNet 3+ [25] | 82.42 | 80.28 |
| | DA [22] | 79.65 | 79.55 | | 83.09 | 81.32 |
| | GloRe [13] | 79.16 | 78.96 | | 82.22 | 79.98 |
| | **Graph-PGCR ($\delta = 2$)** | **81.67** | **79.98** | | **83.56** | **81.63** |
| | **Graph-PGCR ($\delta = 2, 4, 7$)** | **82.03** | **80.95** | | **85.01** | **82.21** |

Table 1. Lung infections segmentation performances on public COVID-19 dataset when gradually adding the proposed components to the UNet.

| Architecture | K | $\delta$ | Dice(%) |
|---|---|---|---|
| Baseline UNet | - | - | 77.50 |
| UNet + DA [22] | - | - | 78.11 |
| UNet + Graph-PGCR (AvgPooling) | 1 | 2 | 79.96 |
| UNet + Graph-PGCR (MaxPooling) | 1 | 2 | 80.02 |
| **UNet + Graph-PGCR (SAP)** | 1 | 2 | 80.84 |
| **UNet + Graph-PGCR (SAP)** | 2 | 2,4 | 81.16 |
| **UNet + Graph-PGCR (SAP)** | 3 | 2,4,7 | 81.38 |

### 3.3. Comparison with the State of the Art

A series of UNet based variations are adopted to further exam the effectiveness of the proposed Graph-PGCR module, including UNet++ [7], UNet-Inf [8] and UNet 3+ [25]. Except for Dual Attention (DA) module, we further compare our method with the state-of-the-art graph context reasoning module (i.e., GloRe) [13]. The hyper-parameters of the graph, e.g., the number of the nodes and its feature dimensions, are set based on [13]. It is worth noting that they are appended at the same place as our proposed module.

(**i**) *Quantitative comparison*: Table 2 shows the comparison results on public and private dataset, where we have the following observations. First, the proposed Graph-PGCR module ($\delta = 2$) improves the performance from the baselines under different segmentation networks. Moreover, our proposed Graph-PGCR module ($\delta = 2$) has superior performance over GloRe module. Additionally, the Graph-PGCR module ($\delta = 2, 4, 7$) with multiple GCR achieves the best performance in four architectures, obtaining average improvement of 3.0 and 2.5 point between four backbones performed on two datasets.

(**ii**) *Qualitative comparison*: Fig.4 visualizes the segmentation results of different plugin based on UNet 3+ network in our private datasets, including DA module, Glore unit, our proposed Graph-PGCR module ($\delta = 2,4,7$). The results illustrated how efficient our proposed Graph-PGCR module is on segmenting the irregular and even small infections. Specifically, it generates segmentation results that are close to the ground truth with much less missegmented infections. The success of Graph-PGCR module is owed to the ability of capturing multi-scale long-range dependencies.

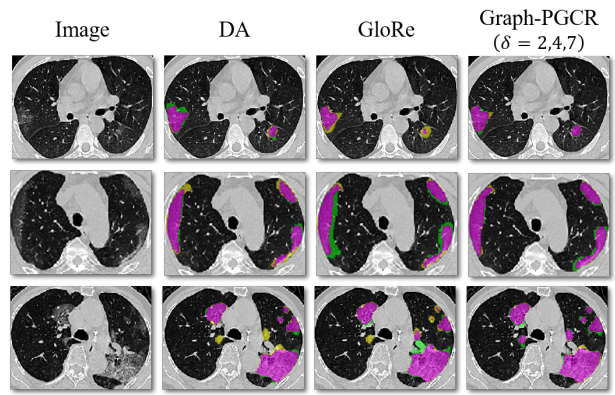

Fig. 4. Qualitative comparisons of different units incorporated with UNet 3+. **Purple areas**: true positive (TP); **Yellow areas**: false negative (FN); **Green areas**: the false positive (FP).

## 4. CONCLUSIONS

In this paper, we develop an effective GCN-based approach, termed as Graph-based Pyramid Global Context Reasoning (Graph-PGCR) module, to model the multi-scale long-range contextual relationships, which is critical for COVID-19 lung infections segmentation. Benefiting from the saliency-aware projection that selects infection-related pixels as graph nodes, a fully-connected graph is constructed where global contextual information is propagated across all nodes via graph convolution. The multi-scale schema is also adopted to explore distinct contextual patterns from multiple graphs. Experiments show that the proposed Graph-PGCR module can effectively capture global contextual dependencies in COVID-19 CT images and consistently improve over four strong baselines on lung infections segmentation task.